\documentclass[conference]{IEEEtran}
\usepackage[T1]{fontenc}
\usepackage{amsmath,bm,amsfonts,amssymb,graphicx,color,mathtools,setspace,
	comment,multirow,xfrac}
\usepackage{booktabs}
\usepackage{epstopdf}
\usepackage{mathtools}
\usepackage{bigints}
\usepackage{cite}
\usepackage{subfigure}
\newcommand*\varhrulefill[1][0.4pt]{\leavevmode\leaders\hrule 
	height#1\hfill\kern0pt}

\begin{document}
	\title{\vspace{-3pt}Hardware Constraints in Compressive Sensing Based Antenna Array\vspace{-5pt}}
	\author{\IEEEauthorblockN{M. Ali Babar Abbasi and 
		Vincent F. Fusco}
	\IEEEauthorblockA{Institute of Electronics, Communications and Information Technology (ECIT), Queen's University Belfast, Belfast, U.K.}
	\IEEEauthorblockA{e--mails: \{m.abbasi~and~v.fusco\}@qub.ac.uk}\vspace{-20pt}}

	\maketitle

	\begin{abstract}
	New constraints based on practical hardware are introduced in compressive sensing (CS) based rectangular antenna array thinning technique. In a standard CS array sparsity enforcement, antenna elements are considered as ideal point sources which do not comply with the practical hardware. It also does not consider the impact of mutual coupling of neighbouring antenna elements on the impedance mismatch. In this work, we propose a combination of constraints based on physical antenna array specifications, mutual coupling and practical antenna element radiation performance in the CS based array thinning enforcement. Analytical modelling along with a design example is presented and discussed. Array performance based on full--wave electromagnetic simulations shows the reliability of the proposed approach.

	\end{abstract}

	\IEEEpeerreviewmaketitle

	\vspace{-1pt}
	\section{Introduction}
	\label{Introduction}
	 
	As per the recent standardization Release 15 by the Third 
	Generation Partnership Project--New Radio (3GPP--NR), millimeter wave (mmWave) carrier frequencies are now an integral part of the 5G mobile communication systems \cite{SHAFI1}. Use of antenna arrays is inevitable for the successful deployment of mmWave 5G infrastructure, especially at the base station (BS) end,  where multi--beam operation with narrow beam width is desirable \cite{HONG1,LARSSON1}. Use of antenna arrays is also a necessity to compensate the electromagnetic (EM) wave's propagation loss, that is very high at the mmWave 5G carrier frequencies compared to the sub--6 GHz spectrum. With a higher number of antenna elements in an array, the associated RF chains are also bound to increase, putting a load on the hardware requirements and deployment cost. Use of thinned and sparse antenna arrays is one of the most prominent techniques used to reduce the number of elements in antenna arrays. This is done by knocking out a group of antenna elements from a two--dimensional (2--D) antenna array aperture by using non--linear optimization methods such as simulated annealing (SA) Genetic algorithm (GA) and compressive sensing (CS). According to CS theory, when certain conditions are considered, some signals can be recovered using fewer measurements compared to traditional methods \cite{CS}. This is applicable to the antenna array design method when we want to get a reference radiation pattern (signal) using fewer number of antenna elements. The resultant thinned array re--adjusts the complex taper and physical locations of antenna elements, while keeping the array projection and scanning capabilities intact. Recent works of using CS for thinned antenna array synthesis focus on the traditional beamforming applications when a far--field radiation pattern is re--created using less antenna elements. The antenna elements are generally considered as point radiating sources, this makes it very difficult to practically achieve the array performance close to the analytical predictions. In addition to this, array comprising of practical antenna elements have physical size constraints. Forcefully decreasing antenna spacing sometimes lead to impractical solution. Recently in \cite{CSIET}, number of novel approaches are proposed which impose the hardware limitations of the antenna arrays, especially at the minimum adjacent antenna separation into the CS. However, when antenna separation is randomized, the real and imaginary component of the mutual impedance becomes unpredictable \cite{mutual} especially when it comes to large antenna arrays. This may lead to degradation in overall array performance, which is highly undesirable when the arrays is to be used for high data--rate communication at the mmWave spectrum.  
	
	In this paper, we propose mutual coupling constraint and practical antenna radiation performance to be included within the CS enforcement in addition to the physical size constraint for array thinning. Section I of the paper describes the problem formulation and analytical model of the method, section II discusses the results while findings are concluded in section III of the paper. Boldface 
	symbols are used to denote matrices and vectors. The transpose,
	Hermititan transpose, and $\ell_n$ norm are denoted by 
	$\left(\cdot\right)^{T}$, 
	$\left(\cdot\right)^{H}$,
	and $\Vert \cdot \Vert_n$ respectively.
	
\vspace{5pt}	

	\section{Proposed Design Method}

		\label{SystemModel}
		
Consider an $M \times N$ uniform rectangular array (URA) distributed along $xy$-plane in the Cartesian coordinate system. The far--field radiation patterns can be defined by \cite{diag}
		
			\begin{equation}
		\label{1}
		f(\theta, \phi) = 
		\sum\limits_{n=0}^{N_{y}-1}
		\sum\limits_{m=0}^{M_{x}-1}
		w_{m,n} e^{jm\frac{2\pi d_x}{\lambda}\sin\theta \cos\phi}
		e^{jn\frac{2\pi d_y}{\lambda}\sin\theta \sin\phi}, 
		\end{equation}		
when $w_{m,n}$ are the complex antenna tappers, $d_x$ and $d_y$ are the adjacent antenna element spacing , and $\lambda$ is the wavelength when $m = 1,...,M-1$ and $n = 1,...,N-1$. Far--field array beam of the array can be represented by

\begin{equation}
	\label{2}
	\textrm{P}(\omega, \theta,\phi) = \mathbf{w}^H \mathbf{S}(\omega,\theta,\phi)
\end{equation}			
when $\mathbf{w}$ is the antenna taper and $\mathbf{S}(\omega,\theta,\phi)$ is the array steering vector, being the function of angular frequency $\omega$, and directions $\theta$ and $\phi$. As discussed in \cite{CSabbasi}, one method of defining a sparse antenna array is to first densely sample the array aperture. This will define a grid of potential antenna locations, when a large number of tapers $\{w_1, w_2,...,w_{M,N}\}$ will be zero. Here, $d_{M-1} \times d_{N-1}$ will define the array aperture size, and $M$ and $N$ will be very large. By finding the optimized number of non--zero \textit{active} antennas to generate the reference $f(\theta,\phi)$, sparsity can be introduced using CS. All the potential antenna locations with $w$ = 0 will be considered inactive, hence can be removed from the array. The CS formulation in this case can be given by

\begin{equation}
\label{3}
\min \Vert\mathbf{w} \Vert_1 \text{ subject to } \Vert\mathbf{p} - \mathbf{w}^H\mathbf{S}\Vert_2 \leq \xi.
\end{equation}			
		
\begin{figure}[!t]
	\centering
	\subfigure[]{\includegraphics[height=3.1 cm]{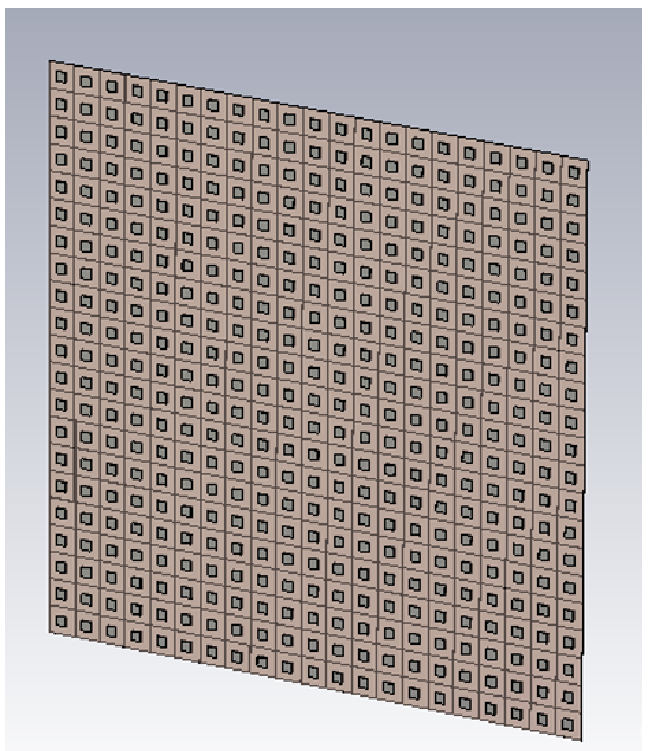}}\hspace{4pt}
	\subfigure[]{\includegraphics[height=3.1 cm]{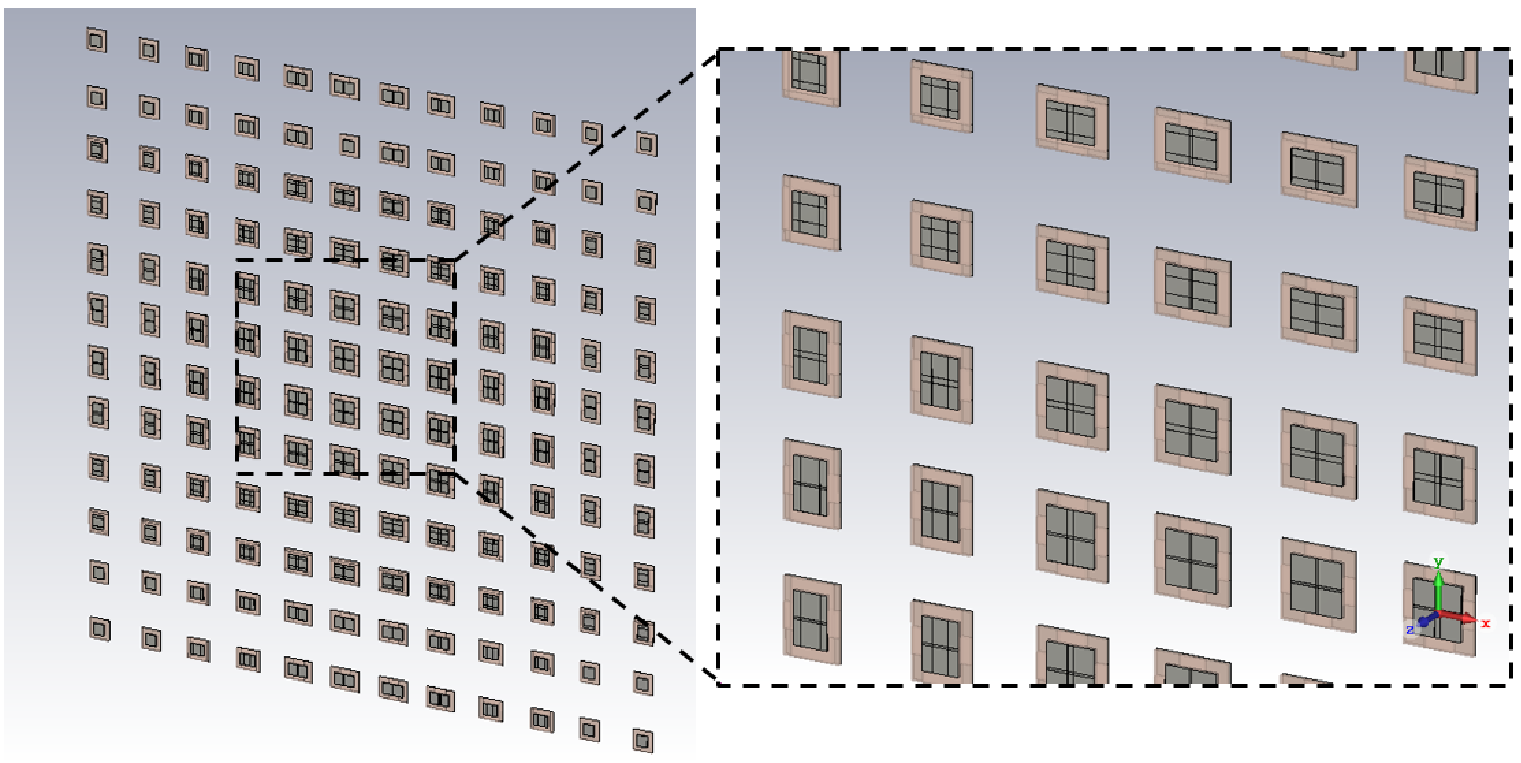}}
	\vspace{-5pt}
	\caption{(a) A conventional 21$\times$21 element microstrip patch antenna URA. (b) Microstrip patch antenna array resulted after using formulation (\ref{3}) when $\xi=10$, representing an impractical solution. }
	\label{imp}
	\vspace{-15pt}
\end{figure}

Here, $\ell_1$ norm is used to approximate $\ell_0$ norm. $\xi$ represents the norm--bounded error margin between designed and the reference $f(\theta,\phi)$ response, also refereed to as CS relaxation factor. The solution in (\ref{3}) is capable of providing antenna locations too close to each other which may not be physically practical. This can be seen from Fig. \ref{imp} where a practical antenna element array was thinned and the $\mathbf{w}$ resulted in forming overlaps of multiple antenna elements, leading to an impractical solution. In a previous studies, we suggested a method of approximating closely spaced antenna elements, excited by a vector summation of the complex tapering of the parent antenna elements \cite{CSabbasi}. In \cite{CSIET}, three methods are introduced to enforce the physical size limitation namely, post--processing closely spaced antenna elements, iterative minimum distance sampling, and re--weighted method. We consider the third method in this study in which $\mathbf{w}$ is converted into re--weighted minimization problem and iterative method is used to find resultant array tapering, then the physical size limitations are enforced. This can be written as

\begin{equation}
\label{4}
\min 
\sum\limits_{k=1}^{K} \gamma_k^i \lvert w_k^i \rvert
\text{; subject to } \Vert\mathbf{p} - \mathbf{w}^H\mathbf{S}\Vert_2 \leq \xi.
\end{equation}
	
when $i$ is the current iteration state, $\gamma_k$ defines the re-weighting terms to solve (\ref{3}) and (\ref{4}) (see section 2.3.3 in \cite{CSIET} for further details). The method here is shown not to always guarantees a viable solution. Moreover, it does not take into consideration the impact of closely placed antenna on other antenna's input impedance. We propose this inclusion as an additional constraint. Consider two closely spaced patch antennas on a substrate with dielectric $\epsilon_{\textrm{r}}$, the mutual impedance is given by

\begin{equation}
\label{5}
Z_{21} = Z_1 - Z_{11} = \frac{1}{I_0^2} \int_V{\vec{E}_{2} \cdotp \vec{J}_1} dV, \end{equation}
where $\vec{E}_{2}$ is the electric field on second antenna due to source current ($\vec{J}_1$) on the first antenna, and $I_0$ is the assumed input current on both antennas. $Z_{11}$ is the self impedance on first antenna while $Z_1$ is the first antenna's impedance when both antennas are simultaneously fed. Similarly, mutual impedance across $M$ and $N$ antenna elements with the antenna at the centre in a URA can be written as  

	\begin{equation}
	\label{6}
	\mathbf{z_{x}} =  
	[Z_{\frac{M-1}{2},1}, Z_{\frac{M-1}{2},2} ,...,Z_{\frac{M-1}{2},M}]^T\text{, and}
	 \end{equation}
	
	\begin{equation}
	\label{7}
	\mathbf{z_{y}} =
	[Z_{\frac{N-1}{2},1}, Z_{\frac{N-1}{2},2} ,...,Z_{\frac{N-1}{2},N}]^T.
	\end{equation}
When $\max(\mathbf{z_{x}})$ and $\max(\mathbf{z_{y}})$ are enforced into the iteration method of formulation (\ref{4}), the resultant tapering $\mathbf{w}$ can provide a solution that considers the physical constraints as well as the mutual impedance impact of the antennas. 
		
\section{Design Example and Discussion}		
		
\begin{figure}[!t]
	\centering
	\subfigure[]{\includegraphics[height=2.5 cm]{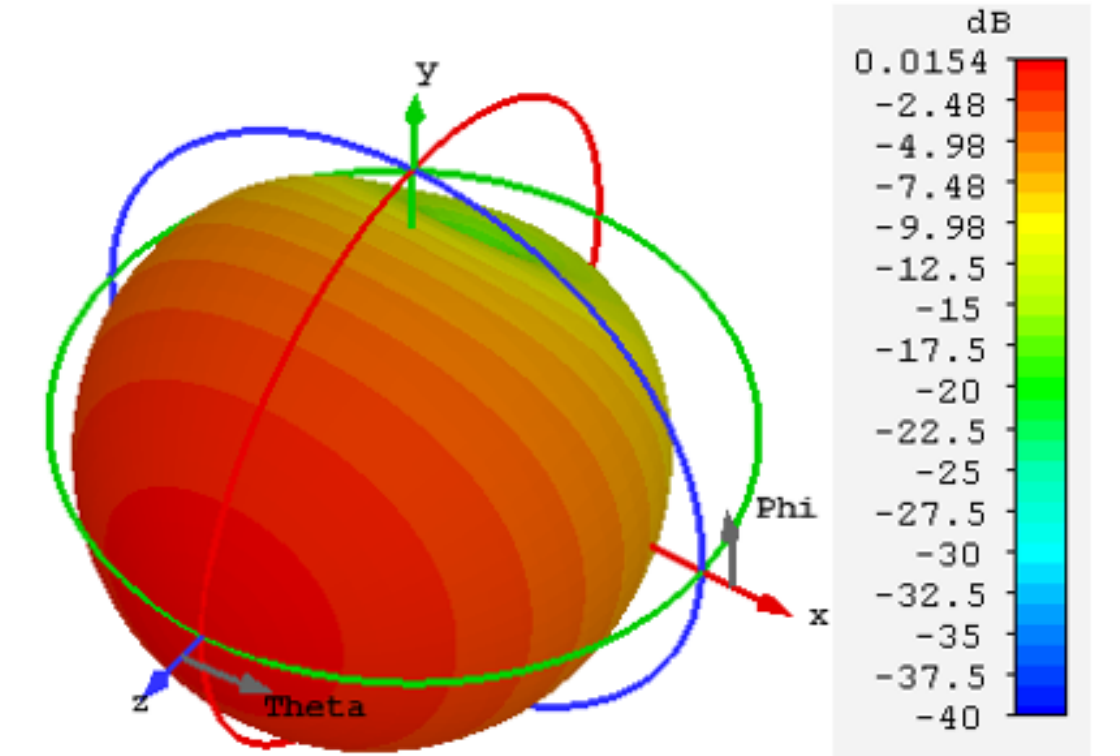}}\hspace{10pt}
	\subfigure[]{\includegraphics[height=2.5 cm]{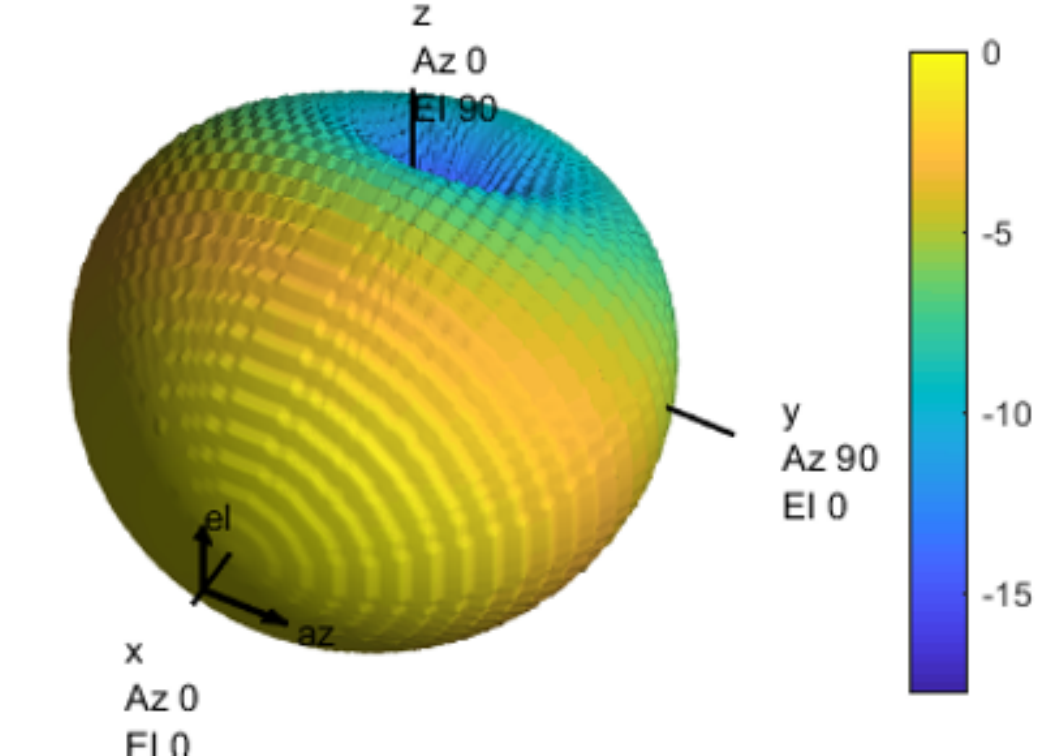}}\hspace{10pt}
	\caption{(a) Far--field normalized patterns numerically evaluated in CST Microwave studio. (b) Imported far--field patterns in Matlab for inclusion in the CS implementation. }
	\label{ff}
	\vspace{-15pt}
\end{figure}

In this design example, we consider a 25$\times$25 element URA of patch antenna elements with a constant $d_x$ and $d_y=\frac{\lambda}{2}$. The array unit cell is designed on 250 $\mu$m thick RO4003 substrate, operating at 28 GHz. 3--dimensional (3--D) far--field patterns of the unit cell is given in Fig. \ref{ff}(a). We first estimate the initial $\mathbf{z_x}$ and $\mathbf{z_y}$ using \cite{pozar}
\vspace{-3pt}
\begin{equation}
\label{8}
S_{12} = \frac{2Z_{12}Z_0}{(Z_{11}+Z_0)(Z_{22}+Z_0)-Z_{12}Z_{21}},
\end{equation}
and used 30 dB Dolph-Chebyshev tapper to define initial $\mathbf{w}$ and weight matrix $\mathbf{W} = [\mathbf{w_1}, \mathbf{w_2},...,\mathbf{w_N}]$. We include the antenna unit--cell patterns estimated from CST Microwave Studio instead of using widely reported ideal isotropic radiating source. It is worth mentioning that the 3-D patterns in CST defined in $\theta$ and $\phi$ were convert to $az$ and $ele$ patterns in Matlab for the computations using the following relations:
\begin{equation}
\label{9}
\sin(el) = \sin\phi\sin\theta \text{, and }
\tan(az) = \cos\phi\tan\theta.
\end{equation}
Converted patterns are shown in Fig. \ref{ff}(b). Matlab was used for the analytical modelling, while CST microwave and design studio co--simulation environment was used for full-wave EM simulations.

\begin{figure}[!t]
	\centering
	\includegraphics[height=3 cm]{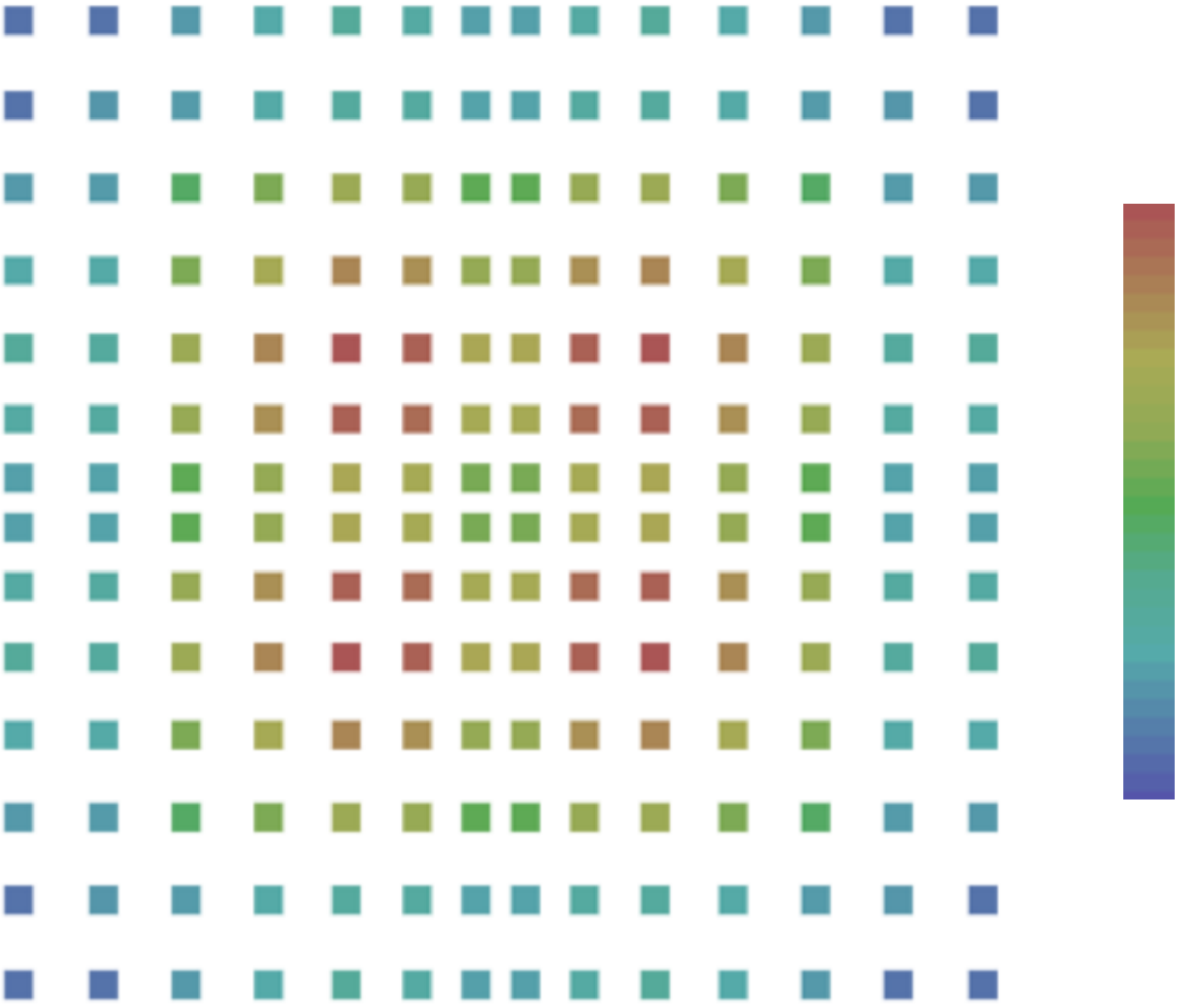}
	\caption{Resultant $\mathbf{W}$ and element location in $xy$-plane after implementing the proposed approach. Colour map represents the absolute magnitude of $\mathbf{W}$ when the map range is from 0 to 14.7.}
	\label{array}
	\vspace{-10pt}
\end{figure}

\begin{figure}[!t]
		\centering
		\subfigure[]{\includegraphics[width=9.1 cm]{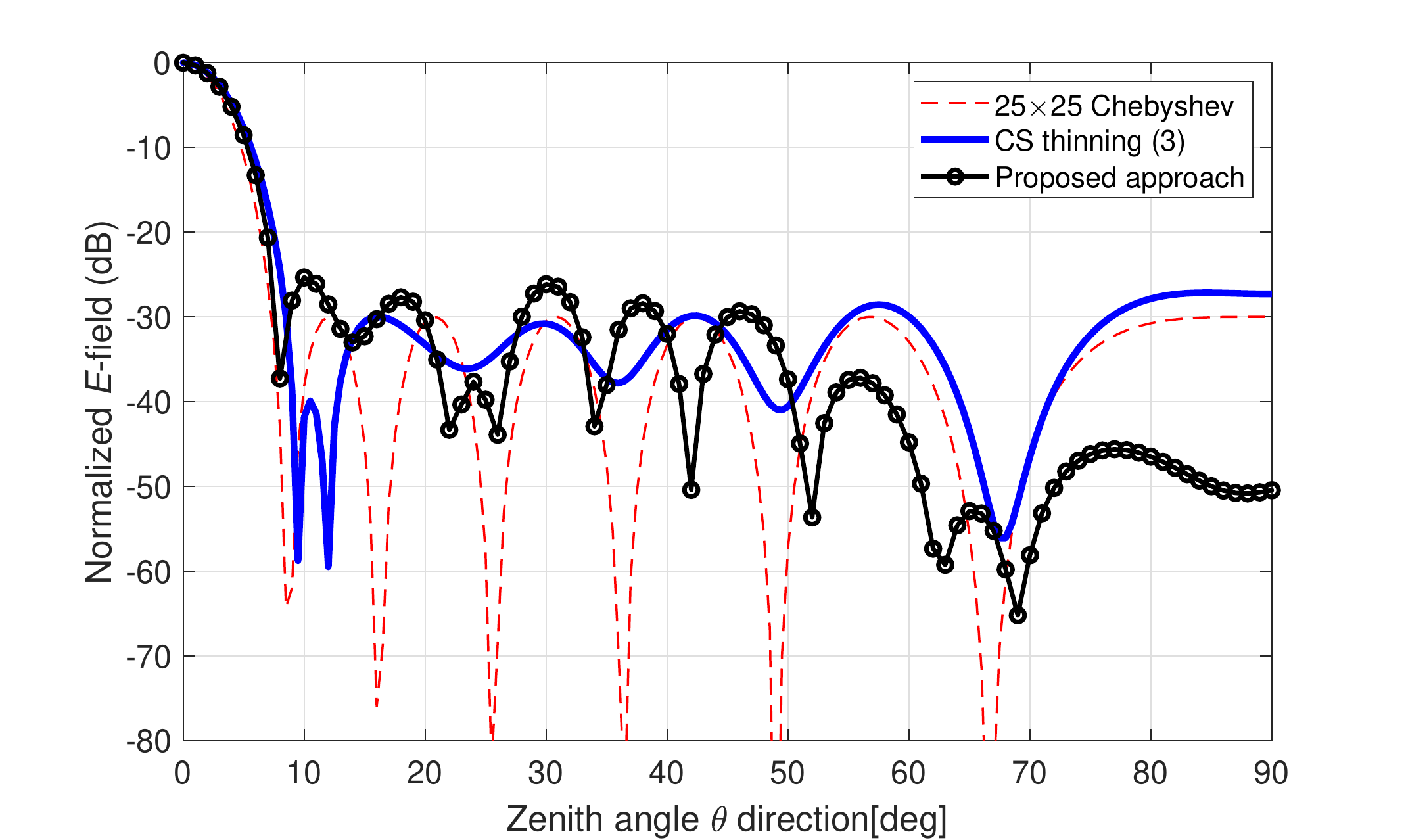}}
		\subfigure[]{\includegraphics[height=3.2 cm]{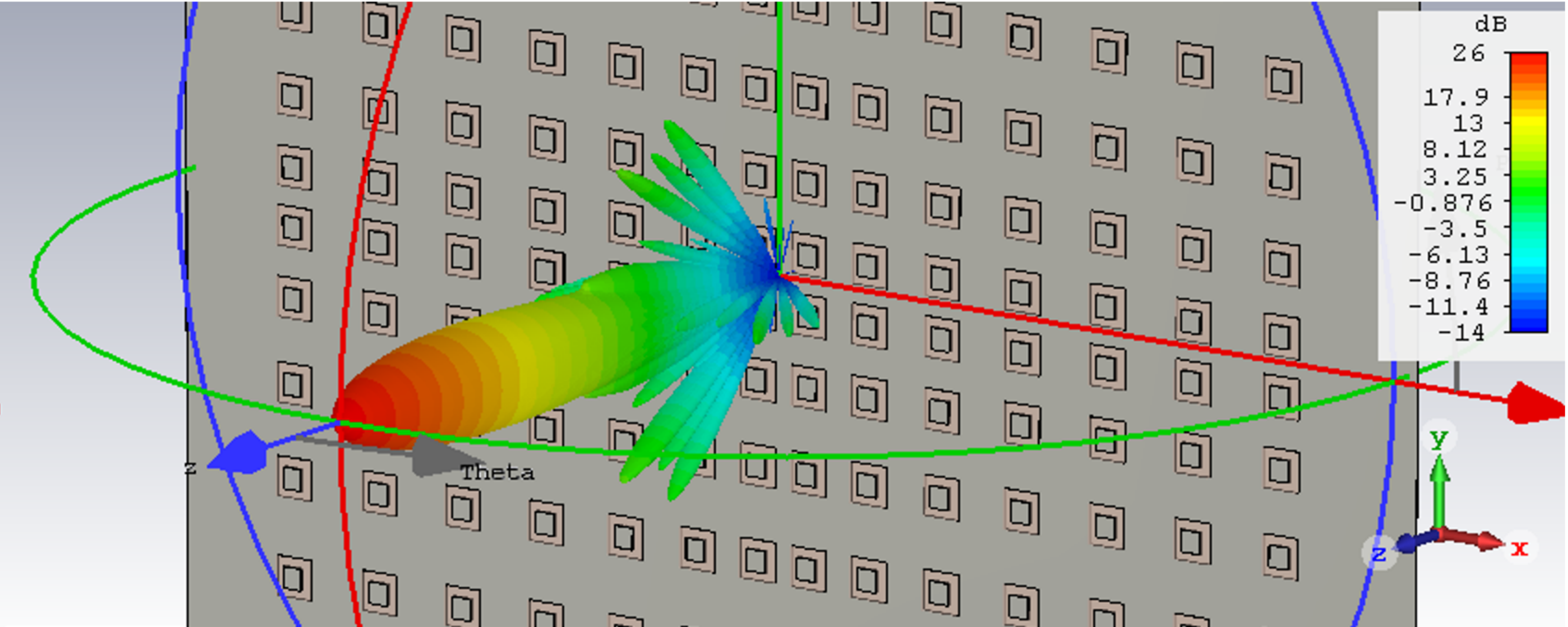}}
		\vspace{-5pt}
		\caption{(a) Far--field patterns comparison between ideal URA, CS implementation considering the physical constrains \cite{CSIET}, and the proposed synthesis approach. We consider $\xi=10$ for the last two graphs. (b) Implementation of the array with patch antennas and far--field absolute realized gain of the array.}
		\label{ff2}

	\end{figure}

Using the approach presented in section II, we achieve array thinning from 625 to 196 elements (i.e. 14$\times$14 elements). The number of total iteration states $i$ was 5 and the maximum number of CS iterations for any given $i$ were 19. This makes the proposed approach computationally extensive compared to previous works. The resultant array has the location of all the antenna elements at a practically realizable location as shown in Fig. \ref{array}. Far--field response of an ideal array with 30 dB Dolph-Chebyshev tapper is given in Fig. \ref{ff2}(a). We use this pattern as a reference for the CS. Re--calibration of the same array using the third approach in \cite{CSIET}, i.e. adding antenna spacing constraint with CS, is also shown for comparison. We use the resultant tapper (Fig. \ref{array}) to estimate far--field response of the practical patch antenna array in CST Microwave Studio. 2--D cut along $yz$-plane of the 3--D patterns is compared in Fig. \ref{ff2}(a) while practically realizable 3--D gain is shown in Fig. \ref{ff2}(b). The 3--dB beam-width of the resultant array is $\sim$0.4$^\circ$ wider compared to the ideal beam of Dolph--Chebyshev array, and almost matches the far--field patterns of the array synthesized using method in \cite{CSIET}. Enforcing mutual coupling reduction and adding physical element constraint cost an increase in the side lobe level up to $\sim$4 dB at around $\pm$10$^\circ$ zenith angle. Side lobe level at the zenith angle > 50$^\circ$ stayed significantly below the targeted -30 dB limit set by the initial Dolph-Chebyshev tapering. A peak gain of 26 dB is achieved from an array using the proposed approach with significantly smaller number of antenna elements compared to conventional array synthesis techniques.

	\section{Conclusion}
	\label{ConclusionsandFutureWorks}
Practically realizable array thinning synthesis approach using Compressive Sensing is presented in this work. Unlike previous works, the approach not only considers the physical placement of the antennas in an array formation, but also considers the mutual impedances, and mismatches caused because of it. In addition to this, practical patch antenna unit cell radiation performance is included in the synthesis process to closely model real--life radiation performance. Other than slight increase in the side--lobes power levels, the synthesized array is shown to match the desired radiation patterns. Future directions include the development of array for 5G base station (BS) at mmWave band with lesser number of RF chains and estimate the cost benefit of the proposed approach.


	\section{Acknowledgements}
	\label{Acknowledgment}
This work was supported by the EPSRC, U.K., under grant EP/P000673/1 and grant EP/EN02039/1. Thansk to E. Candes, J. Romberg
(Caltech), S. Boyd (Stanford) for public dissemination of convex optimization code. Thanks to U. Naeem for discussion around Compressive Sensing. 
	
	\bibliographystyle{IEEEtran}

\end{document}